\documentclass[aps,prl,reprint,article,showpacs]{revtex4-1}
\usepackage{blindtext}
\usepackage{graphicx}
\usepackage{amsfonts}
\usepackage{amsmath}

\begin{document}

\title{Compressed sensing for Hamiltonian reconstruction}
\author{Kenneth Rudinger}
\email{Current address:  Sandia National Laboratories, Albuquerque, NM 87185-1322.\\Email:  kmrudin@sandia.gov}
%\affiliation{Sandia National Laboratories, Albuquerque, New Mexico 87123}
\author{Robert Joynt}
\affiliation{University of Wisconsin-Madison, Madison, Wisconsin 53706}

\begin{abstract}
In engineered quantum systems, the Hamiltonian is often not completely known and needs to be
determined experimentally with accuracy and efficiency. We show that this may be done at
temperatures that are greater than the characteristic interaction energies, but not too much greater.
The condition for this is that there are not too many interactions: the Hamiltonian is sparse in a well-defined sense.
The protocol that accomplishes this is related to compressed sensing methods of classical signal processing;
in this case applied to sparse rather than low-rank matrices.\end{abstract}


\maketitle

\section{Introduction}

In quantum physics, the standard method for understanding a large system has
long been to make an approximate model Hamiltonian that captures the
essential physics of the material in question. \ More recently, this
situation is often turned on its head - a quantum system of $n$ qubits is
constructed and we need to find its Hamiltonian from experimental data. \ To
do quantum information processing of any kind, accurate control of the
Hamiltonian is always a prerequisite. \ One needs to be able to apply
external controls to guide the desired time-dependent Hamiltonian, but it is
usually also the case that there are \textquotedblleft always-on" terms,
generally time-independent or nearly so, in the Hamiltonian that need to be
determined at a precise quantitative level \cite{Shulman2014}. \ This is a particularly
pressing issue for a quantum memory, or in cold-atom systems that are
specifically constructed in order to simulate many-body Hamiltonians. \ For
electron spin qubits in semiconductor quantum dots \cite{DiVincenzo:1998p419}%
, for example, the single-qubit energy-level splittings are subject to
unknown random hyperfine fields, and there are also dipole-dipole
interactions. \ These are one- and two-qubit interactions, but there is also
the more challenging case of multi-qubit interactions. \ \ In this paper we
propose an efficient way to determine these \textquotedblleft always-on"\
terms. \ 

For $n=1$ and $n=2,$ considerable work has been done, since these cases are
relevant to the performance of gates \cite{Cole:2005p062312,
Devitt:2006p052317, Schirmer:2009p022333}. \ Process tomography is the usual
tool for problems with $n>2$, but standard methods \cite%
{NielsenBook,Chuang1997} require a number of measurements that scales
exponentially with $n.$ \ Other methods that pertain particularly to spin
systems require only a small number of measurements, but they appear to
involve full simulation of the system, a task that again scales
exponentially \cite{Burgarth:2009p020305,
BurgarthNJP:2009,BurgarthNJP:2011,DiFranco:2009p187203}. \ Several authors
have investigated the use of techniques from compressed sensing \cite%
{Candes2008} which would give an efficient solution to this problem when the
process matrix $\chi$ is $s-$sparse (has only $s$ nonzero elements) in some
basis \cite{Shabani:2011p100401, Baldwin2014, Rodionov2014}. \ The number of measurements
needed to determine $\chi $ is then $O\left( sn\right) .$ However, this
scheme requires prior knowledge of the basis in which $\chi $ is sparse.
Thus it is useful for verifying quantum gates, but cannot be used to
determine entirely unknown processes (or Hamiltonians), which is the case we
are considering.

As pointed out in Ref.~\cite{Shabani:2011p012107}, it makes sense to take
advantage of the fact that, to a very good approximation, almost all qubit
Hamiltonians $H$ have only one- and two-qubit interactions, so that the
number of parameters to be determined scales only as $n^{2}.$ \ These
authors suggest a sequence of randomly chosen measurements on randomly
prepared states. \ If the time interval $t$ between preparation and
measurement is short enough: \ $\left\vert \left\vert H\right\vert
\right\vert ~t<<1$, then the density matrix is simply related to $H.$ \ Here 
$\left\vert \left\vert H\right\vert \right\vert $ is the operator norm
(largest eigenvalue) of $H.$ \ Compressed-sensing techniques can then come
into play and the number of measurements required to determine $H$ is $%
O(n^{3}).$ \ However, $\left\vert \left\vert H\right\vert \right\vert $
grows with the size of the system, which limits the usefulness of this
scheme. \ 

\section{Method}

Here we propose a different approach for the experimental determination of $%
H.$ \ The most general Hamiltonian for an array of $n$ qubits is:

\begin{equation}
H=-\eta \sum_{a=1}^{4^{n}-1}J_{a}\lambda _{a}
\end{equation}%
where $a$ is an $n$-digit base-4 number $a=a_{1}a_{2}...a_{n}$ and the $%
\lambda _{a}$ are tensor products of Pauli matrices: $\lambda _{a}=\sigma
_{a_{1}}\otimes \sigma _{a_{2}}\otimes ...\otimes \sigma _{a_{n}}$. \ $%
\sigma _{1,2,3}=\sigma _{x,y,z}$ and $\sigma _{0}$ is the identity matrix. \
For notational convenience we have defined the energy scale $\eta $, set by
the condition that the dimensionless variables $J_{a}$ satisfy $\left\vert
J_{a}\right\vert \leq 1.$ We will assume that only $s$ of the $4^{n}-1$
possible $J_{a}$ are zero and $s<<4^{n}$. \ The system is placed in a bath
and comes to thermal equilibrium. \ The density matrix is $\rho =\exp \left(
-\beta H\right) /Q,$ where $Q$ is the partition function: $Q=$ Tr $\exp
\left( -\beta H\right) $ and $\beta =1/k_{B}T$. \ If $T=0,$ $\rho $ reduces
to $\rho =\left\vert 0\right\rangle \left\langle 0\right\vert $ where $%
\left\vert 0\right\rangle $ is the ground state so that the density matrix
has rank 1. \ We will work in the opposite, high-temperature, limit$\ \eta
\beta <<1,$ where  $\ \rho =I-\beta H+\beta ^{2}H^{2}/2+....$ and we may
truncate the expansion. \ In general there are a macroscopic number of
energy eigenstates that enter $\rho $ and $\rho $ represents a high rank
state. \ It is important to note that the application of compressed sensing
proposed here is opposite to others in the literature that primarily focus
on the determination of states of low rank \ \cite%
{Gross:2010p150401,Flammia2012}. \ In fact the density matrix is technically
of \textit{full} rank at any finite temperature and the naive (but
inefficient) procedure to determine the $J_{a}$ would be to measure the
observables $\lambda _{a}$. \ For $\eta \beta <<1$ this gives $\eta
J_{a}=-2^{-n}~$Tr~$\left( \lambda _{a}H\right) \approx %\left( 2^{-n}/\beta
%\right)
\beta ^{-1}~$Tr$~\left( \lambda _{a}\rho \right) .$ \ However, most of the
diagonal matrix elements are exponentially small, and we will use this fact
to reduce the number of measurements that need to be made.

The measurement and processing protocol is as follows. After the system
reaches equilibrium, its state is given by $\rho
=2^{-n}I+2^{-n}\sum_{a=1}^{4^{n}-1}v_{a}\lambda _{a},$ where $\vec{v}$ is
the equilibrium polarization vector of the system. We then subject the
system to a random unitary transformation $U$ so that the new state of the
system is $\rho ^{\prime }=U\rho U^{-1}$. The procedure for generating
random $U$'s that are efficiently implementable with a small gate set is a
modification of one proposed for quantum data hiding by DiVincenzo, Leung,
and Terhal \cite{DiVincenzo2001}, using work by Harrow and Low on random
quantum circuits \cite{Harrow2008}. \ The $U$'s are not selected uniformly
from the Haar distribution but our results indicate that they provide usable
compression matrices. (Details for generating each $U$ are provided in Appendix A)

The new polarization vector $\vec{v}^{\prime }$ is linearly related to the
previous one: $v_{a}^{\prime }=$\ $\sum_{b=1}^{4^{n}-1}C_{ab}v_{b}$ with $%
C_{ab}=2^{-n}$ Tr $\left( \lambda _{a}U\lambda _{b}U^{-1}\right) .$ \ $C$ is
an orthogonal matrix and $\vec{v}$ is a long but approximately sparse
vector, the ``signal vector". \ We now measure $M$ of the observables $%
\lambda $ obtaining the results $\left\{ y_{k}\right\} _{k=1}^{M}$ with the $%
y_{k}$ satisfying $-1\leq y_{k}\leq 1.$ \ We will discuss the magnitude of $%
M $ and the choice of the $\lambda $'s below. \ $\vec{y}$ is our
``measurement vector", a subset of the elements of $\vec{v}$. \ We now have 
\begin{equation}
y_{k}=\sum\limits_{b}C_{kb}^{\left( M\right) }v_{b},
\end{equation}%
where $C^{\left( M\right) }$ consists of $M$ rows of $C,$ the choice of rows
corresponding to the measurements taken. \ $C^{\left( M\right) }$ is an $%
M\times \left( 4^{n}-1\right) $ matrix, the ``compression matrix". \ The
next step is to estimate the polarization vector by minimizing the $L_{1}$
norm of all possible polarization vectors that are consistent with the
measurement results:%
\begin{equation}
\vec{v}_{est}=\arg \min_{\vec{w}} \left\vert \left\vert \vec{w}\right\vert
\right\vert _{1},~\text{subject to }\sum\limits_{b}C_{kb}^{\left( M\right)
}w_{b}=y_{k}.
\end{equation}%
The $L_{1}$ norm of a vector $\vec{w}$ is defined as $\left\vert \left\vert 
\vec{w}\right\vert \right\vert _{1}=\sum\limits_{i=1}^{d}\left\vert
w_{i}\right\vert .$ \ This is a convex optimization problem that can be
solved efficiently. \ For our purposes it is important to note that this
compressed sensing protocol is stable with respect to deviations from exact
sparsity in the signal vector, so that, as we shall see below, the protocol
works at moderate temperatures. \ Also, it can be shown that if $C^{\left(
M\right) }$ is formed by choosing rows at random from $C,$ then $C^{\left(
M\right) }$ satisfies a certain restricted isometry condition which
guarantees that that if $M>A$ $n~\ln ^{3}s$ we can recover $\vec{v}$ with
high probability, Here $A$ is a constant. \cite{Vershynin2010}.

Once a good estimate of the polarization vector is available, we can
estimate the Hamiltonian: 
\begin{equation}
H_{est}=\beta ^{-1}(2^{-n}\text{Tr}\left( \ln \rho _{est}\right) ~I-\ln \rho
_{est}).  \label{eq:hest}
\end{equation}

\section{Results}

We now turn to numerical studies of the protocol for 3, 4 and 5 qubits, for
which $a$ takes on $N=63$, $N=255$, and $N=1023$ values, respectively. \ We
input a random Hamiltonian, compute the equilibrium density matrix $\rho $,
and perform $M$ measurements, i.e., characterize $\rho $ by the numbers Tr$%
\left( \lambda _{i}\rho \right) ,$ $i=1,2,...,M.$ \ (This is our definition
of a measurement; while experimental measurements in a lab are subject to finite sample
error, it is known that compressed sensing is robust against such errors \cite
{Candes2008}.)
While measurements are chosen at random, they are ordered
by weight, that is all measurements of weight one (i.e., single-qubit
measurements) are performed before all measurements of weight two (i.e.,
two-qubit measurements), and so on. (See Appendix B for further explanation.)

The simplest case is the determination of the $J_{a}$ when we are given that
only $s$ of them are nonzero. \ We do not have firm guarantees of success at
finite temperature, since the density matrix is not $s$-sparse. \ So the
first task is to determine how high the temperature needs to be to ensure
success. \ The temperature is quantified by the dimensionless ratio $\eta
\beta .$ \ Success is measured by the distance of $H_{est},$ the Hamiltonian
estimated from Eq. \ref{eq:hest}, from the actual Hamiltonian $H,$ the
metric chosen as the one corresponding to the Frobenius norm: if $%
(\left\vert \left\vert H_{est}-H\right\vert \right\vert_F )/\eta <$
threshold, the procedure is judged to have succeeded$.$

Fig. 1 shows the quality of the reconstruction of $H$ as a function of the
parameters $M/N$, which is the number of measurements divided by the signal
length, and the sparsity ratio $s/N.$ \ There are 3 qubits and each pixel in
the plots is the result of 100 trials. \ Note first that the lower right
corner is a region where the number of nonzero entries in $J_{a}$ is greater
than the number of measurements: reconstruction is impossible there. \ As we
move away from the diagonal to the upper left, the success probability
increases. \ As is generally observed in cases where compressed sensing
works, the boundary between success and failure (that is, the Donoho-Tanner phase
transition) is sharp. High temperature is favorable for reconstruction, but
even at quite moderate temperatures there is a very substantial region of
parameter space where the determination of $H$ succeeds. \ The red region in
both panels is where $H$ is successfully reconstructed, due to the density
matrix being approximately sparse in that region.

\begin{figure}[tbp]
\begin{center}
\includegraphics[width = 1\linewidth]{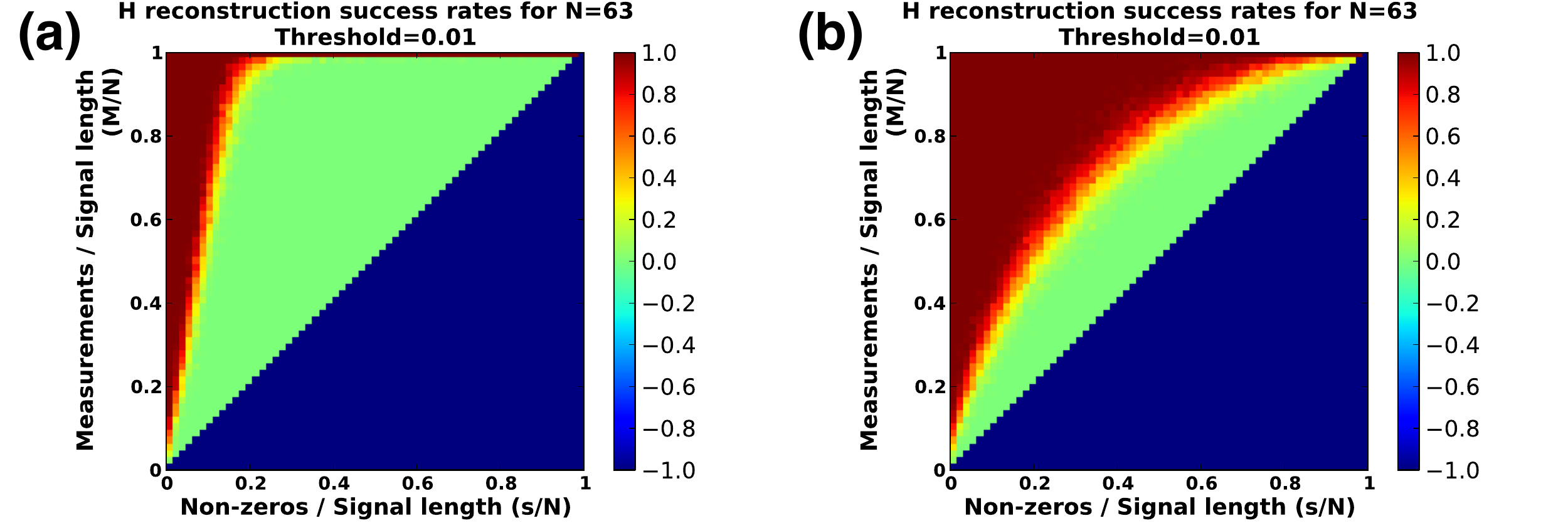}
\end{center}
\caption[Hamiltonian reconstruction heat maps for 3 qubits ]{Quality of
Hamiltonian determination for random couplings as a function of temperature.
\ In (a) and (b) the inverse dimensionless temperature is given by $\protect%
\eta \protect\beta =10^{-4}$ and $\protect\eta \protect\beta =10^{-1}$
respectively. \ Red indicates a high success rate, green indicates failure,
and a\ ``negative'' success rate (blue) means that reconstruction is impossible.
Each pixel is an average of 100 trials. \ }
\label{fig:HeatMaps}
\end{figure}

These computations show that compressed sensing can work in principle, and
gives strong evidence that the number of measurements needed is proportional
to $n,$ the number of qubits, rather than $N,$ the number of possible
couplings, when the Hamiltonian is sparse. \ However, equipped with the
knowledge that $H$ is sparse, quantum state tomography can also be carried
out with a reduced number of meaurements. \ We next examine the question of
how much advantage is actually gained in practice over the straightforward
method of standard tomography, stopping when $H$ has been determined. \ Fig.
2 gives this comparison for $n=3$ [Fig. 2(a)], $n=4$ [Fig. 2(b)], and $n=5$
[Fig. 2(c)], with small values of $s,$ and for a moderate temperature of $%
\eta \beta =10^{-1}.$ \ The number of trials per data point is 100. The
sampled $M$'s have a spacing of 1 for $n=3$ and $n=4$, starting at a value
of $M=2$; due to computational constraints, every tenth value of $M$ is used
for $n=5$, starting at a value of $M=11$. \ The median value of the
normalized quality $(\left\vert \left\vert H_{est}-H\right\vert
\right\vert_F )/\left\vert \left\vert H\right\vert \right\vert_F $ of the
estimate is plotted as a function of $M,$ so that low values correspond to
accurate estimates. \ When the curve drops off sharply, the ``phase
transition" from failure to success has occurred. \ Thus for example, in
Fig. 2(a), the compressed sensing (CS) protocol for $n=3$ and $s=1$ succeeds
at $M=5.$ \ It is seen that compressed sensing gives a large saving in the
number of measurements for all cases considered, ranging (roughly) from a
factor of 4 to 7 for $n=3$,
from 6 to 12 for $n=4$, and from 12 to as high as 50 for $n=5$. 
This is good evidence that the advantage of the compressed sensing protocol
increases with $n$, as we would expect from the scaling arguments above.

\begin{figure}[h!]
\begin{center}
\includegraphics[width = 1 \linewidth]{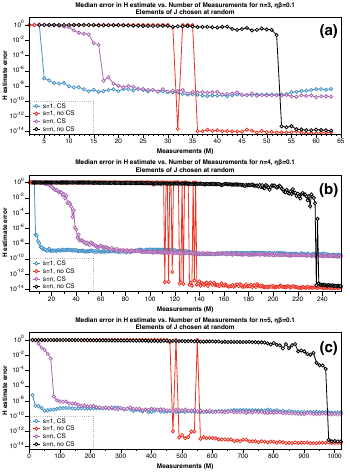}
\end{center}
\par
\vspace{-0.5 cm} 
\caption[Quality of Hamiltonian reconstruction with and without compressed
sensing, for random couplings]{Quality of Hamiltonian determination for
random couplings as a function of number of qubits. (a), (b), and (c) give
the error in the estimated Hamiltonian for $n=3$, 4 and $5$ qubits,
respectively, as a function of the number of measurements made ($M$), with
compressed sensing (CS) and without (no CS).  [While the minimum error obtained
by the no CS protocol appears to be lower than the minimum error obtained by the CS
protocol ($\sim10^{-14}$ and $\sim10^{-9}$, respectively), these differences are
simply artifacts of different noise floors of different numerical methods.  For each algorithm, achieving
its respective noise floor indicates that the Hamiltonian has been successfully reconstructed.]
In each case, the CS protocol substantially decreases the $M$ required to accurately reconstruct the Hamiltonian. 
\ The improvement increases with the number of qubits. Each data point is the median value of 100 trials.}
\label{fig:Scatter1}
\end{figure}

In most cases of actual physical interest, we not only have some knowledge
of the sparsity of $H,$ we also have some knowledge of where the nonzeros
lie. \ For example, for spin qubits, 1- and 2-body interactions are likely
to be much greater in magnitude than 3- and higher-body interactions. \ We
then find $s=O\left( n^{2}\right) .$ \ Locality may also reduce the
sparsity; for sufficiently short-range interactions $s=O\left( n\right) $. \
This is a very different situation than we have considered so far, where the
nonzero $J_{a}$ were taken at random. \ Of course exponential reductions in $M$
required to reconstruct $H$ are now out of the question. \ The question is whether we can still get
speedups that may be useful in real situations - even constant speedups can
be important. \ So we perform the same numerical experiment as in Fig. 2,
but now the nonzero $J_{a}$ are restricted to those corresponding to $%
\lambda _{a}$ that are 1- and 2-qubit operators, \textit{i.e.}, $a$ has at
most 2 nonzero digits. \ The results are shown in Fig. 3. \ The number of
trials and all other parameters are the same as in Fig. 2. In the ``no CS''
(standard tomography) protocol, measurements of 1- and 2-body operators are
made first, which now improves the performance of the ``no CS'' procedure,
but not enough to overcome the advantage of the CS\ protocol
(see Appendix C for further details).

\ The ratio of the number of measurements required is about a factor of 2 to
4 for $n=3$, about a factor of 3 to 6 for $n=4$, and about a factor of 6 to
8 for $n=5$. \ Thus the speedup is less when the knowledge of the locations
of the nonzeros is increased, but it is still quite substantial. \ More
importantly, it appears that the speedup still increases with the number of
qubits. 
\begin{figure}[h!]
\begin{center}
\includegraphics[width = 1 \linewidth]{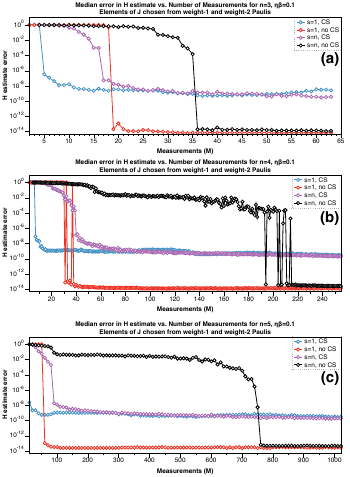}
\end{center}
\par
\vspace{-0.5 cm}
\caption[Quality of Hamiltonian reconstruction with and without compressed
sensing, for 1- and 2-qubit couplings]{\ Quality of Hamiltonian
determination for 1- and 2-qubit couplings as a function of number of
qubits. (a), (b), and (c) give the error in the estimated Hamiltonian for $%
n=3$, 4, and 5 qubits, respectively, as a function of the number of
measurements made ($M$), with compressed sensing (CS) and without (no CS).  
(As in Figure 2, the minimum error obtained by the two different protocols is
an artifact of the different noise floors for the numerical methods employed; 
in each cases, achieving the noise floor indicates successful Hamiltonian
reconstruction.)  In each case, the CS protocol substantially decreases the $M$
required to accurately reconstruct the Hamiltonian, though this improvement is not as
big as for random couplings. \ As before, the improvement increases as the number of qubits
increases. Each data point is the median value of 100 trials.}
\label{fig:Scatter2}
\end{figure}
\section{Conclusion}
Previous improvements in efficiency of quantum state tomography have shown
the usefulness of compressed sensing techniques by focusing on the
reconstruction of states of low rank. This work, by contrast, uses this
technique to reconstruct states of high rank. This is not useful for
validation of gate quality, but it can be used to determine the parameters
in a many-body Hamiltonian.

As compressed sensing reduces the number of real-valued system parameters that must be measured, 
at the cost of increased post-processing, compressed sensing is only of value for systems in which measurements are
expensive but signal processing and post-processing are cheap.
his tradeoff is highly attractive for many classical applications, but the
tradeoffs vary from case to case.   Our quantum protocol will be useful and attractive when 
measurement settings are expensive, but quantum gate operations are cheap.
Otherwise, straightforward tomography will be better. The competition
between the two is greatly affected by how much advance knowledge we have
about the system. It is when we do not have a very good idea in advance
about the shape of the Hamiltonian that our method is useful.

\subsection{Acknowledgments}

We thank John Aidun, Eric Bach, Charles Baldwin, Robin Blume-Kohout, S. N. Coppersmith, Daniel Crow, Adam Frees, Mark Friesen, John Gamble, Kevin Jamieson, Amir Kalev, and Robert Nowak for useful discussions.

\section{Appendices}

\subsection{A. Generation of $U$}

To choose a random unitary map that is efficiently implementable with a
small gate set, we use the following procedure, inspired by a technique for
quantum data hiding proposed by DiVincenzo, Leung, and Terhal \cite%
{DiVincenzo2001}, along with work by Harrow and Low on random quantum
circuits \cite{Harrow2008}.

For an $n$-qubit system, we consider the following set $\mathcal{G}$ of
quantum gates

\begin{equation}
\mathcal{G} = \{H_p,P_q, P_r^\dagger, R_s\left(\tfrac{\pi}{8}\right),
CNOT_{tu}\},
\end{equation}
where $H$ is the Hadamard gate, $P$ is the phase gate, $R\left(\tfrac{\pi}{8}%
\right)$ is the $\tfrac{\pi}{8}$ gate, and $CNOT$ is the controlled-not
gate. The subscripts label the qubit (or qubits) that each gate is acting
on, that is, $\mathcal{G}$ contains all single-qubit copies of $%
\{H,P,P^\dagger, R\left(\tfrac{\pi}{8}\right)\}$ and all two-qubit copies of 
$CNOT$.

To form the unitary map $U$, we simply select (with replacement) $n^8$
elements of $\mathcal{G}$ uniformly at random. Letting $g_i$ denote the $%
i^{th}$ selection from $\mathcal{G}$, we define $U$ to be given by

\begin{equation}
U = \prod_{i=1}^{n^8} g_i.
\end{equation}

Note that this gives us a random unitary operation on $n$ qubits which,
while not selected uniformly from the Haar distribution, is sufficiently
random as to successfully generate a compression matrix which can be used
for compressed sensing. Additionally, we note that it is an open question as
to whether or not a smaller set of gates and/or a shorter gate sequence
could yield equally successful results. 
%\cite{DiVincenzo:1998p419} \cite{Cole:2005p062312}

\subsection{B.  Weight-ordering of measurements}
It may be of some benefit to the experimentalist for whom lower weight measurements
are easier to perform to be able to prioritize low-weight measurements over
high-weight measurements.

Therefore, we show here that %, from a compressed sensing perspective,
the order the measurements are chosen in should not affect
the accuracy of the Hamiltonian or the density matrix reconstructions, allowing 
for the measurements to be chosen according to weight.  (That is to say, all single-qubit measurements
may performed before any two-qubit measurement, which in turn may precede all three-qubit 
measurements, and so on.)
%Paulis may be chosen according to weight order without affecting 
%weight-ordering 
%While each measurement $\lambda_i$ is chosen at random, the order in which they 
%are chosen is determined by weight-ordering.  (That is to say, all single-qubit measurements
%are performed before any two-bit measurement, which in turn precede all three-qubit 
%measurements, and so on.)  %This may reduce the difficulty of the task for the experimentalist
%As exper
This ordering by weight is justified in the following manner.  %While each measurement 
%$\lambda_i$ is chosen at random from the set of all 
%Pauli measurements (excluding those already measured), the
%Suppose we have order, at random, ordering on 

We note that if the $k^{th}$ Pauli measured is $\lambda_k$, then the $k^{th}$ element of
our measurement vector $\vec{y}$ is given as
\begin{equation}
\label{eq:y_k2}
y_k = \text{Tr}\left(\lambda_k U^\dagger \rho U \right),
\end{equation}
where $\rho$ is the initial density matrix and $U$ is the random unitary map.  However,
due to the cyclic property of the trace, we may re-express Eq.~\eqref{eq:y_k2} as

\begin{equation}
y_k = \text{Tr}\left( \left(U\lambda_k U^\dagger\right) \rho \right).
\end{equation}
That is, we may consider our $k^{th}$ measurement to correspond to measuring
the expectation value a Pauli subjected to a random unitary transformation with
respect to the fixed and original density matrix.  Therefore, as $U$ effectively 
randomizes each $\lambda_k$, choosing them in order of their weights should
not affect the reconstruction algorithm.  (Indeed, we have performed numerical 
tests which demonstrate this.)

\subsection{C.  State reconstruction via ``no CS'' protocol}
The ``no CS'' protocol for reconstructing the density matrix $\rho$ is as follows. For
an estimate of $\rho$ in which $M$ Pauli measurements are allowed, the $M$
expectation values are input as the appropriate $v_i$'s; the remaining $v_i$%
's are set to zero. While this estimation procedure could theoretically
produce a non-physical $\rho_{est}$ with negative eigenvalues, in practice
this is not a concern as any state we are estimating has a polarization
vector with a small $L_2$ norm, while a non-physical density matrix with
one or more negative eigenvalues has a polarization vector with a large $L_2$
norm.

\bibliographystyle{apsrev4-1}
\bibliography{citations}

\begin{thebibliography}{21}%
\makeatletter
\providecommand \@ifxundefined [1]{%
 \@ifx{#1\undefined}
}%
\providecommand \@ifnum [1]{%
 \ifnum #1\expandafter \@firstoftwo
 \else \expandafter \@secondoftwo
 \fi
}%
\providecommand \@ifx [1]{%
 \ifx #1\expandafter \@firstoftwo
 \else \expandafter \@secondoftwo
 \fi
}%
\providecommand \natexlab [1]{#1}%
\providecommand \enquote  [1]{``#1''}%
\providecommand \bibnamefont  [1]{#1}%
\providecommand \bibfnamefont [1]{#1}%
\providecommand \citenamefont [1]{#1}%
\providecommand \href@noop [0]{\@secondoftwo}%
\providecommand \href [0]{\begingroup \@sanitize@url \@href}%
\providecommand \@href[1]{\@@startlink{#1}\@@href}%
\providecommand \@@href[1]{\endgroup#1\@@endlink}%
\providecommand \@sanitize@url [0]{\catcode `\\12\catcode `\$12\catcode
  `\&12\catcode `\#12\catcode `\^12\catcode `\_12\catcode `\%12\relax}%
\providecommand \@@startlink[1]{}%
\providecommand \@@endlink[0]{}%
\providecommand \url  [0]{\begingroup\@sanitize@url \@url }%
\providecommand \@url [1]{\endgroup\@href {#1}{\urlprefix }}%
\providecommand \urlprefix  [0]{URL }%
\providecommand \Eprint [0]{\href }%
\providecommand \doibase [0]{http://dx.doi.org/}%
\providecommand \selectlanguage [0]{\@gobble}%
\providecommand \bibinfo  [0]{\@secondoftwo}%
\providecommand \bibfield  [0]{\@secondoftwo}%
\providecommand \translation [1]{[#1]}%
\providecommand \BibitemOpen [0]{}%
\providecommand \bibitemStop [0]{}%
\providecommand \bibitemNoStop [0]{.\EOS\space}%
\providecommand \EOS [0]{\spacefactor3000\relax}%
\providecommand \BibitemShut  [1]{\csname bibitem#1\endcsname}%
\let\auto@bib@innerbib\@empty
%</preamble>
\bibitem [{\citenamefont {Shulman}\ \emph {et~al.}(2014)\citenamefont
  {Shulman}, \citenamefont {Harvey}, \citenamefont {Nichol}, \citenamefont
  {Bartlett}, \citenamefont {Doherty}, \citenamefont {Umansky},\ and\
  \citenamefont {Yacoby}}]{Shulman2014}%
  \BibitemOpen
  \bibfield  {author} {\bibinfo {author} {\bibfnamefont {M.~D.}\ \bibnamefont
  {Shulman}}, \bibinfo {author} {\bibfnamefont {S.~P.}\ \bibnamefont {Harvey}},
  \bibinfo {author} {\bibfnamefont {J.~M.}\ \bibnamefont {Nichol}}, \bibinfo
  {author} {\bibfnamefont {S.~D.}\ \bibnamefont {Bartlett}}, \bibinfo {author}
  {\bibfnamefont {A.~C.}\ \bibnamefont {Doherty}}, \bibinfo {author}
  {\bibfnamefont {V.}~\bibnamefont {Umansky}}, \ and\ \bibinfo {author}
  {\bibfnamefont {A.}~\bibnamefont {Yacoby}},\ }\href@noop {} {\  (\bibinfo
  {year} {2014})},\ \Eprint {http://arxiv.org/abs/1405.0485v1}
  {arXiv:1405.0485v1 [cond-mat]} \BibitemShut {NoStop}%
\bibitem [{\citenamefont {DiVincenzo}\ and\ \citenamefont
  {Loss}(1998)}]{DiVincenzo:1998p419}%
  \BibitemOpen
  \bibfield  {author} {\bibinfo {author} {\bibfnamefont {D.~P.}\ \bibnamefont
  {DiVincenzo}}\ and\ \bibinfo {author} {\bibfnamefont {D.}~\bibnamefont
  {Loss}},\ }\href@noop {} {\bibfield  {journal} {\bibinfo  {journal}
  {Superlatt Microstruct}\ }\textbf {\bibinfo {volume} {23}},\ \bibinfo {pages}
  {419 } (\bibinfo {year} {1998})}\BibitemShut {NoStop}%
\bibitem [{\citenamefont {Cole}\ \emph {et~al.}(2005)\citenamefont {Cole},
  \citenamefont {Schirmer}, \citenamefont {Greentree}, \citenamefont {Wellard},
  \citenamefont {Oi},\ and\ \citenamefont {Hollenberg}}]{Cole:2005p062312}%
  \BibitemOpen
  \bibfield  {author} {\bibinfo {author} {\bibfnamefont {J.~H.}\ \bibnamefont
  {Cole}}, \bibinfo {author} {\bibfnamefont {S.~G.}\ \bibnamefont {Schirmer}},
  \bibinfo {author} {\bibfnamefont {A.~D.}\ \bibnamefont {Greentree}}, \bibinfo
  {author} {\bibfnamefont {C.~J.}\ \bibnamefont {Wellard}}, \bibinfo {author}
  {\bibfnamefont {D.~K.~L.}\ \bibnamefont {Oi}}, \ and\ \bibinfo {author}
  {\bibfnamefont {L.~C.~L.}\ \bibnamefont {Hollenberg}},\ }\href {\doibase
  10.1103/PhysRevA.71.062312} {\bibfield  {journal} {\bibinfo  {journal} {Phys.
  Rev. A}\ }\textbf {\bibinfo {volume} {71}},\ \bibinfo {pages} {062312}
  (\bibinfo {year} {2005})}\BibitemShut {NoStop}%
\bibitem [{\citenamefont {Devitt}\ \emph {et~al.}(2006)\citenamefont {Devitt},
  \citenamefont {Cole},\ and\ \citenamefont {Hollenberg}}]{Devitt:2006p052317}%
  \BibitemOpen
  \bibfield  {author} {\bibinfo {author} {\bibfnamefont {S.~J.}\ \bibnamefont
  {Devitt}}, \bibinfo {author} {\bibfnamefont {J.~H.}\ \bibnamefont {Cole}}, \
  and\ \bibinfo {author} {\bibfnamefont {L.~C.~L.}\ \bibnamefont
  {Hollenberg}},\ }\href {\doibase 10.1103/PhysRevA.73.052317} {\bibfield
  {journal} {\bibinfo  {journal} {Phys. Rev. A}\ }\textbf {\bibinfo {volume}
  {73}},\ \bibinfo {pages} {052317} (\bibinfo {year} {2006})}\BibitemShut
  {NoStop}%
\bibitem [{\citenamefont {Schirmer}\ and\ \citenamefont
  {Oi}(2009)}]{Schirmer:2009p022333}%
  \BibitemOpen
  \bibfield  {author} {\bibinfo {author} {\bibfnamefont {S.~G.}\ \bibnamefont
  {Schirmer}}\ and\ \bibinfo {author} {\bibfnamefont {D.~K.~L.}\ \bibnamefont
  {Oi}},\ }\href {\doibase 10.1103/PhysRevA.80.022333} {\bibfield  {journal}
  {\bibinfo  {journal} {Phys. Rev. A}\ }\textbf {\bibinfo {volume} {80}},\
  \bibinfo {pages} {022333} (\bibinfo {year} {2009})}\BibitemShut {NoStop}%
\bibitem [{\citenamefont {Nielsen}\ and\ \citenamefont
  {Chuang}(2000)}]{NielsenBook}%
  \BibitemOpen
  \bibfield  {author} {\bibinfo {author} {\bibfnamefont {M.~A.}\ \bibnamefont
  {Nielsen}}\ and\ \bibinfo {author} {\bibfnamefont {I.~L.}\ \bibnamefont
  {Chuang}},\ }\href@noop {} {\emph {\bibinfo {title} {Quantum Computation and
  Quantum Information}}}\ (\bibinfo  {publisher} {Cambridge University Press},\
  \bibinfo {address} {Cambridge},\ \bibinfo {year} {2000})\BibitemShut
  {NoStop}%
\bibitem [{\citenamefont {{Chuang}}\ and\ \citenamefont
  {{Nielsen}}(1997)}]{Chuang1997}%
  \BibitemOpen
  \bibfield  {author} {\bibinfo {author} {\bibfnamefont {I.~L.}\ \bibnamefont
  {{Chuang}}}\ and\ \bibinfo {author} {\bibfnamefont {M.~A.}\ \bibnamefont
  {{Nielsen}}},\ }\href {\doibase 10.1080/09500349708231894} {\bibfield
  {journal} {\bibinfo  {journal} {Journal of Modern Optics}\ }\textbf {\bibinfo
  {volume} {44}},\ \bibinfo {pages} {2455} (\bibinfo {year}
  {1997})}\BibitemShut {NoStop}%
\bibitem [{\citenamefont {Burgarth}\ \emph {et~al.}(2009)\citenamefont
  {Burgarth}, \citenamefont {Maruyama},\ and\ \citenamefont
  {Nori}}]{Burgarth:2009p020305}%
  \BibitemOpen
  \bibfield  {author} {\bibinfo {author} {\bibfnamefont {D.}~\bibnamefont
  {Burgarth}}, \bibinfo {author} {\bibfnamefont {K.}~\bibnamefont {Maruyama}},
  \ and\ \bibinfo {author} {\bibfnamefont {F.}~\bibnamefont {Nori}},\ }\href
  {\doibase 10.1103/PhysRevA.79.020305} {\bibfield  {journal} {\bibinfo
  {journal} {Phys. Rev. A}\ }\textbf {\bibinfo {volume} {79}},\ \bibinfo
  {pages} {020305} (\bibinfo {year} {2009})}\BibitemShut {NoStop}%
\bibitem [{\citenamefont {Burgarth}\ and\ \citenamefont
  {Maruyama}(2009)}]{BurgarthNJP:2009}%
  \BibitemOpen
  \bibfield  {author} {\bibinfo {author} {\bibfnamefont {D.}~\bibnamefont
  {Burgarth}}\ and\ \bibinfo {author} {\bibfnamefont {K.}~\bibnamefont
  {Maruyama}},\ }\href@noop {} {\bibfield  {journal} {\bibinfo  {journal} {New.
  J. Phys.}\ }\textbf {\bibinfo {volume} {11}},\ \bibinfo {pages} {103019}
  (\bibinfo {year} {2009})}\BibitemShut {NoStop}%
\bibitem [{\citenamefont {Burgarth}\ \emph {et~al.}(2011)\citenamefont
  {Burgarth}, \citenamefont {Maruyama},\ and\ \citenamefont
  {Nori}}]{BurgarthNJP:2011}%
  \BibitemOpen
  \bibfield  {author} {\bibinfo {author} {\bibfnamefont {D.}~\bibnamefont
  {Burgarth}}, \bibinfo {author} {\bibfnamefont {K.}~\bibnamefont {Maruyama}},
  \ and\ \bibinfo {author} {\bibfnamefont {F.}~\bibnamefont {Nori}},\
  }\href@noop {} {\bibfield  {journal} {\bibinfo  {journal} {New. J. Phys.}\
  }\textbf {\bibinfo {volume} {13}},\ \bibinfo {pages} {013019} (\bibinfo
  {year} {2011})}\BibitemShut {NoStop}%
\bibitem [{\citenamefont {Di~Franco}\ \emph {et~al.}(2009)\citenamefont
  {Di~Franco}, \citenamefont {Paternostro},\ and\ \citenamefont
  {Kim}}]{DiFranco:2009p187203}%
  \BibitemOpen
  \bibfield  {author} {\bibinfo {author} {\bibfnamefont {C.}~\bibnamefont
  {Di~Franco}}, \bibinfo {author} {\bibfnamefont {M.}~\bibnamefont
  {Paternostro}}, \ and\ \bibinfo {author} {\bibfnamefont {M.~S.}\ \bibnamefont
  {Kim}},\ }\href {\doibase 10.1103/PhysRevLett.102.187203} {\bibfield
  {journal} {\bibinfo  {journal} {Phys. Rev. Lett.}\ }\textbf {\bibinfo
  {volume} {102}},\ \bibinfo {pages} {187203} (\bibinfo {year}
  {2009})}\BibitemShut {NoStop}%
\bibitem [{\citenamefont {Cand\`{e}s}\ and\ \citenamefont
  {Wakin}(2008)}]{Candes2008}%
  \BibitemOpen
  \bibfield  {author} {\bibinfo {author} {\bibfnamefont {E.}~\bibnamefont
  {Cand\`{e}s}}\ and\ \bibinfo {author} {\bibfnamefont {M.}~\bibnamefont
  {Wakin}},\ }\href {\doibase 10.1109/MSP.2007.914731} {\bibfield  {journal}
  {\bibinfo  {journal} {Signal Processing Magazine, IEEE}\ }\textbf {\bibinfo
  {volume} {25}},\ \bibinfo {pages} {21} (\bibinfo {year} {2008})}\BibitemShut
  {NoStop}%
\bibitem [{\citenamefont {Shabani}\ \emph
  {et~al.}(2011{\natexlab{a}})\citenamefont {Shabani}, \citenamefont {Kosut},
  \citenamefont {Mohseni}, \citenamefont {Rabitz}, \citenamefont {Broome},
  \citenamefont {Almeida}, \citenamefont {Fedrizzi},\ and\ \citenamefont
  {White}}]{Shabani:2011p100401}%
  \BibitemOpen
  \bibfield  {author} {\bibinfo {author} {\bibfnamefont {A.}~\bibnamefont
  {Shabani}}, \bibinfo {author} {\bibfnamefont {R.~L.}\ \bibnamefont {Kosut}},
  \bibinfo {author} {\bibfnamefont {M.}~\bibnamefont {Mohseni}}, \bibinfo
  {author} {\bibfnamefont {H.}~\bibnamefont {Rabitz}}, \bibinfo {author}
  {\bibfnamefont {M.~A.}\ \bibnamefont {Broome}}, \bibinfo {author}
  {\bibfnamefont {M.~P.}\ \bibnamefont {Almeida}}, \bibinfo {author}
  {\bibfnamefont {A.}~\bibnamefont {Fedrizzi}}, \ and\ \bibinfo {author}
  {\bibfnamefont {A.~G.}\ \bibnamefont {White}},\ }\href {\doibase
  10.1103/PhysRevLett.106.100401} {\bibfield  {journal} {\bibinfo  {journal}
  {Phys. Rev. Lett.}\ }\textbf {\bibinfo {volume} {106}},\ \bibinfo {pages}
  {100401} (\bibinfo {year} {2011}{\natexlab{a}})}\BibitemShut {NoStop}%
\bibitem [{\citenamefont {Baldwin}\ \emph {et~al.}(2014)\citenamefont
  {Baldwin}, \citenamefont {Kalev},\ and\ \citenamefont
  {Deutsch}}]{Baldwin2014}%
  \BibitemOpen
  \bibfield  {author} {\bibinfo {author} {\bibfnamefont {C.}~\bibnamefont
  {Baldwin}}, \bibinfo {author} {\bibfnamefont {A.}~\bibnamefont {Kalev}}, \
  and\ \bibinfo {author} {\bibfnamefont {I.}~\bibnamefont {Deutsch}},\
  }\href@noop {} {\  (\bibinfo {year} {2014})},\ \Eprint
  {http://arxiv.org/abs/1404.2877v2} {arXiv:1404.2877v2 [quant-ph]}
  \BibitemShut {NoStop}%
\bibitem [{\citenamefont {Rodionov}\ \emph {et~al.}(2014)\citenamefont
  {Rodionov}, \citenamefont {Veitia}, \citenamefont {Barends}, \citenamefont
  {Kelly}, \citenamefont {Sank}, \citenamefont {Wenner}, \citenamefont
  {Martinis}, \citenamefont {Kosut},\ and\ \citenamefont
  {Korotkov}}]{Rodionov2014}%
  \BibitemOpen
  \bibfield  {author} {\bibinfo {author} {\bibfnamefont {A.~V.}\ \bibnamefont
  {Rodionov}}, \bibinfo {author} {\bibfnamefont {A.}~\bibnamefont {Veitia}},
  \bibinfo {author} {\bibfnamefont {R.}~\bibnamefont {Barends}}, \bibinfo
  {author} {\bibfnamefont {J.}~\bibnamefont {Kelly}}, \bibinfo {author}
  {\bibfnamefont {D.}~\bibnamefont {Sank}}, \bibinfo {author} {\bibfnamefont
  {J.}~\bibnamefont {Wenner}}, \bibinfo {author} {\bibfnamefont {J.~M.}\
  \bibnamefont {Martinis}}, \bibinfo {author} {\bibfnamefont {R.~L.}\
  \bibnamefont {Kosut}}, \ and\ \bibinfo {author} {\bibfnamefont {A.~N.}\
  \bibnamefont {Korotkov}},\ }\href@noop {} {\  (\bibinfo {year} {2014})},\
  \Eprint {http://arxiv.org/abs/1407.0761} {arXiv:1407.0761 [quant-ph]}
  \BibitemShut {NoStop}%
\bibitem [{\citenamefont {Shabani}\ \emph
  {et~al.}(2011{\natexlab{b}})\citenamefont {Shabani}, \citenamefont {Mohseni},
  \citenamefont {Lloyd}, \citenamefont {Kosut},\ and\ \citenamefont
  {Rabitz}}]{Shabani:2011p012107}%
  \BibitemOpen
  \bibfield  {author} {\bibinfo {author} {\bibfnamefont {A.}~\bibnamefont
  {Shabani}}, \bibinfo {author} {\bibfnamefont {M.}~\bibnamefont {Mohseni}},
  \bibinfo {author} {\bibfnamefont {S.}~\bibnamefont {Lloyd}}, \bibinfo
  {author} {\bibfnamefont {R.~L.}\ \bibnamefont {Kosut}}, \ and\ \bibinfo
  {author} {\bibfnamefont {H.}~\bibnamefont {Rabitz}},\ }\href {\doibase
  10.1103/PhysRevA.84.012107} {\bibfield  {journal} {\bibinfo  {journal} {Phys.
  Rev. A}\ }\textbf {\bibinfo {volume} {84}},\ \bibinfo {pages} {012107}
  (\bibinfo {year} {2011}{\natexlab{b}})}\BibitemShut {NoStop}%
\bibitem [{\citenamefont {Gross}\ \emph {et~al.}(2010)\citenamefont {Gross},
  \citenamefont {Liu}, \citenamefont {Flammia}, \citenamefont {Becker},\ and\
  \citenamefont {Eisert}}]{Gross:2010p150401}%
  \BibitemOpen
  \bibfield  {author} {\bibinfo {author} {\bibfnamefont {D.}~\bibnamefont
  {Gross}}, \bibinfo {author} {\bibfnamefont {Y.-K.}\ \bibnamefont {Liu}},
  \bibinfo {author} {\bibfnamefont {S.~T.}\ \bibnamefont {Flammia}}, \bibinfo
  {author} {\bibfnamefont {S.}~\bibnamefont {Becker}}, \ and\ \bibinfo {author}
  {\bibfnamefont {J.}~\bibnamefont {Eisert}},\ }\href {\doibase
  10.1103/PhysRevLett.105.150401} {\bibfield  {journal} {\bibinfo  {journal}
  {Phys. Rev. Lett.}\ }\textbf {\bibinfo {volume} {105}},\ \bibinfo {pages}
  {150401} (\bibinfo {year} {2010})}\BibitemShut {NoStop}%
\bibitem [{\citenamefont {Flammia}\ \emph {et~al.}(2012)\citenamefont
  {Flammia}, \citenamefont {Gross}, \citenamefont {Liu},\ and\ \citenamefont
  {Eisert}}]{Flammia2012}%
  \BibitemOpen
  \bibfield  {author} {\bibinfo {author} {\bibfnamefont {S.~T.}\ \bibnamefont
  {Flammia}}, \bibinfo {author} {\bibfnamefont {D.}~\bibnamefont {Gross}},
  \bibinfo {author} {\bibfnamefont {Y.-K.}\ \bibnamefont {Liu}}, \ and\
  \bibinfo {author} {\bibfnamefont {J.}~\bibnamefont {Eisert}},\ }\href
  {http://stacks.iop.org/1367-2630/14/i=9/a=095022} {\bibfield  {journal}
  {\bibinfo  {journal} {New Journal of Physics}\ }\textbf {\bibinfo {volume}
  {14}},\ \bibinfo {pages} {095022} (\bibinfo {year} {2012})}\BibitemShut
  {NoStop}%
\bibitem [{\citenamefont {DiVincenzo}\ \emph {et~al.}(2001)\citenamefont
  {DiVincenzo}, \citenamefont {Leung},\ and\ \citenamefont
  {Terhal}}]{DiVincenzo2001}%
  \BibitemOpen
  \bibfield  {author} {\bibinfo {author} {\bibfnamefont {D.~P.}\ \bibnamefont
  {DiVincenzo}}, \bibinfo {author} {\bibfnamefont {D.~W.}\ \bibnamefont
  {Leung}}, \ and\ \bibinfo {author} {\bibfnamefont {B.~M.}\ \bibnamefont
  {Terhal}},\ }\href@noop {} {\  (\bibinfo {year} {2001})},\ \Eprint
  {http://arxiv.org/abs/quant-ph/0103098v1} {arXiv:quant-ph/0103098v1
  [quant-ph]} \BibitemShut {NoStop}%
\bibitem [{\citenamefont {Harrow}\ and\ \citenamefont
  {Low}(2009)}]{Harrow2008}%
  \BibitemOpen
  \bibfield  {author} {\bibinfo {author} {\bibfnamefont {A.~W.}\ \bibnamefont
  {Harrow}}\ and\ \bibinfo {author} {\bibfnamefont {R.~A.}\ \bibnamefont
  {Low}},\ }\href@noop {} {\  (\bibinfo {year} {2009})},\ \Eprint
  {http://arxiv.org/abs/0802.1919v3} {arXiv:0802.1919v3 [quant-ph]}
  \BibitemShut {NoStop}%
\bibitem [{\citenamefont {Vershynin}(2010)}]{Vershynin2010}%
  \BibitemOpen
  \bibfield  {author} {\bibinfo {author} {\bibfnamefont {R.}~\bibnamefont
  {Vershynin}},\ }\href@noop {} {\  (\bibinfo {year} {2010})},\ \Eprint
  {http://arxiv.org/abs/1011.3027v7} {arXiv:1011.3027v7 [math.PR]} \BibitemShut
  {NoStop}%
\end{thebibliography}%

\end{document}